\documentclass[manuscript]{aastex}

\usepackage[utf8]{inputenc}

\shorttitle{CCD photometry of bright stars using objective wire mesh}
\shortauthors{Kamiński et al.}

\begin{document}

\title{CCD photometry of bright stars using objective wire mesh}

\author{Krzysztof Kamiński}
\affil{Astronomical Observatory Institute, Faculty of Physics, A. Mickiewicz
University, Słoneczna 36, 60-286 Poznań, Poland.}
\email{chrisk@amu.edu.pl}

\author{Aleksander Schwarzenberg-Czerny}
\affil{Copernicus Astronomical Centre, ul. Bartycka 18, PL 00-716
Warsaw, Poland}

\and

\author{Marika Zgórz}
\affil{Astronomical Observatory Institute, Faculty of Physics, A. Mickiewicz
University, Słoneczna 36, 60-286 Poznań, Poland.}

\begin{abstract}
Obtaining accurate photometry of bright stars from the ground remains tricky because 
of the danger of overexposure of the target and/or lack of suitable nearby comparison 
star. The century-old method of the objective wire mesh used to produce 
multiple stellar images seems attractive for 
precision CCD photometry of such stars. Our tests 
on $\beta$ Cep and its comparison star differing by 5 magnitudes prove very
encouraging. Using a CCD camera
and a 20 cm telescope with objective covered
with a plastic wire mesh, located in poor weather conditions we obtained differential 
photometry of precision 4.5 mmag per 2 min exposure.
Our technique is flexible and may be tuned to cover as big 
magnitude range as 6 -- 8 magnitudes. We discuss the possibility of installing 
a wire mesh directly in the filter wheel.
\end{abstract}


\section{Introduction} \label{s0}
Renewed interest in studies of bright stars in general stems from their
suitability to long term spectroscopic monitoring with modest telescopes for
asteroseismic purposes. As a byproduct of the extra-solar planet quest emerged
the new generation of fiber-fed echelle spectrographs capable of measuring radial
velocities of those stars accurate to meters per second. This opens a new
window for studies of multiple/low amplitude coherent and stochastic (solar-type)
oscillations of luminous stars. Both kinds of oscillations are of great use for
asteroseismology, particularly to constrain the efficiency of convection and mixing
in stellar interiors. However, precise mode identification demands knowledge
of phase shifts between velocity and light curves, as well as color dependence
of photometric amplitudes \citep{das02}.

Accurate photometry of bright stars remains tricky because of danger of
overexposure of the target and/or lack of suitable nearby comparison stars. 
Last half century produced relatively few long-term light curves for such stars.
It may be argued that the best results can be obtained from Space and using wide
angle cameras. This became the motivation for the constellation of BRITE
nano-satellites in the process of launching (Orleański et al., 2010). In the
present paper we investigate suitability of a venerable photographic technique
to obtain good quality CCD photometry of very bright stars {\em from the
ground}. For this purpose we applied a CCD camera fitted to a 20cm
telescope with its objective covered with a dense wire mesh.

Late XIX century attempts by astronomers to employ photographic plates for
stellar photometry were hampered by the need to calibrate a non-linear response
of the photographic emulsion to light. For {\em Carte du Ciel}  Kapteyn in 1891
proposed to make alternate exposures with and without
wire mesh cover of the objective to vary aperture (c.f. \citealt{wea46}). 
Later, \citet{her10} noted that a sufficiently dense wire mesh would produce
multiple diffraction images for each star,
thus alleviating the need for multiple exposures. He argued
that the rate of illumination between different images of the same star would
remain fixed. His idea applied either for the direct images of the sky or for
images of the calibration source exposed on the edge of the plate. 
However, as far as we are aware, no images of sky taken through the wire
mesh were reported in the electronic age of astronomy.

\section{Methods} \label{s1}
\subsection{Instrument setup}\label{s11}
We employed two small instruments. First was the 10cm f/5 guide telescope on top of 
the 0.7m Alt-Az Poznan Spectroscopic Telescope 2, fitted with SBIG ST-7 camera 
(hereafter PST2G, see www.astro.amu.edu.pl/GATS for general reference). 
The metal mesh with $0.1 \times 0.1$ mm pitch and $0.06$ mm wire width
was fitted on its V filter. 
Second was a 20cm f/4.4 Orion Optics Newtonian telescope on 
a Celestron CGE Pro equatorial mount, equipped with SBIG 
ST-8 camera (hereafter Orion). The plastic mesh with $1.5 \times 1.5$ mm pitch
and $0.5$ mm wire width was fitted on its objective.
Both telescopes were located in Poznań University Observatory 
park, 65m above sea level in downtown Poznań, a city of 0.5 mln inhabitants. 
The local astro-climate is mediocre at best, affected by 
the surrounding city, with unstable extinction, 
often significantly different between western and eastern sky.

The purpose of the mesh was to produce multiple 
stellar images so that the 1-st or 2-nd order diffraction images of bright stars 
became properly exposed while their 0-order images remain overexposed on purpose, 
to reveal 0-order images of comparison stars at a comparable S/N level.
Diffraction of light of the bright stars on the wire mesh produced multiple 
diffraction images roughly separated by 1.9 and 1.2 arcmin respectively 
for PST2G and Orion, at the image scale of 3.71 and 2.11 arcsec/pixel (Fig. \ref{fig1}). 
The perpendicular wires and corresponding diffraction patterns are not needed 
for our purposes, except that they ensure mechanical stiffness of the mesh 
and allow wider selection of n-th order diffraction images.
Orion observations were made through R filter and exposed for 150s, 
PST2G observations were made through V filter and exposed for 10s.
A rotation of the mesh was introduced for the charge bleeding
from saturated pixels of the central image not to interfere with
the diffraction orders of choice.

\subsection{Diffraction physics} \label{s12}
Let us consider a grating of N parallel wires of diameter $\epsilon$ separated 
by distance d. Their diffraction pattern corresponds to that of N slits of 
width $\delta=d-\epsilon$, 
\begin{equation}
I_1(x)=\frac{I_0}{N^2}\left[\frac{\sin(\pi x\delta/\lambda)}{
\pi x\delta/\lambda}\frac{\sin(N\pi xd/\lambda)}{
\sin(\pi xd/\lambda)}\right]^2\label{eq1}
\end{equation}
where $I_0$ denotes intensity observed without grating (i.e. for $\delta=d$), 
$x$ is the angle with respect to the normal to the grating and $\lambda$ is 
wavelength
(e.g. \citealt{cra68}). Maxima (fringes) are observed when second denominator 
vanishes, i.e. when $x$ satisfies $xd/\lambda=m$, where $m$ is an integer. 
So the angular separation of fringes in radians is 
\begin{equation}
\Delta x_1=\frac{\lambda}{d}.\label{eq2}
\end{equation}

For narrow slits, $\delta\ll d$, 
the first factor remains close to $1$ and all maxima appear of comparable 
height. On the contrary, for thin wires, $\epsilon=d-\delta\ll d$, 
all fringes except the central one are fainter by a factor 
of $(\epsilon/d)^2$. This is so since the fringe maxima in first 
factor $\sin^2(\pi x\delta/\lambda)=\sin^2(\pi 
m-\pi x\epsilon/\lambda)\approx(\pi x\epsilon/\lambda)^2$ and the whole factor 
becomes $(\epsilon/d)^2$. However, for the central fringe, $x=0$, 
first factor remains equal to 1.

For a mesh of perpendicular wires the fringing pattern becomes the product of 
those in $x$ and $y$ directions, i.e.

\begin{equation}
I(x,y)=\frac{I_1(x)I_1(y)}{I_0}.\label{eq3}
\end{equation}

In such case the non-central fringes compared to the central one are fainter 
by a factor of $(\epsilon/d)^4$. Thus the target-comparison star dynamic range 
of our method may be increased by thinning of wires. From Eq. (\ref{eq1}) it
follows that the diffraction fringes are quite narrow, $\Delta x_2\approx 
\frac{1}{Nd}=\frac{1}{D}$, corresponding 
to the diffraction on the whole aperture of diameter $D$. In fact, 
it can be demonstrated that the fringe pattern corresponds to the squared 
absolute value of 2D Fourier transform of the aperture pattern. 
In particular the first and second factors in Eq. (\ref{eq1}) correspond 
respectively to Fourier transforms of a single slit of width $\delta$ 
and N slits of width 0. Any distortions and 
asymmetries of the individual fringes observed in Fig. \ref{fig1} are 
consequence of long-range deformations of the wire mesh. However, 
the observed image constitutes convolution of the diffraction pattern 
with the seeing profile, hence the actual diameter of the low-order fringes 
is determined by seeing. The high-order fringes constitute grating spectra. 

In one dimension a flat grid represented by a real function $f(x)$ yields 
a symmetric diffraction pattern $P(-X)=|{\cal F}f(-X)|^2= 
|\overline{{\cal F}f(+X)}|^2=P(X)$, where $x, X$ 
denote grid and image plane coordinates. Therefore, 
the diffraction asymmetry observed in Fig. 1 requires for grid 
function $f$ to have an imaginary part, corresponding to phase difference 
of the incoming plane light wave falling on different sections of the grid. 
This happens for the grid tilted/warped out of the flat objective 
plane perpendicular to the optical axis, say by several light wavelengths. 

\section{Results} \label{s2}
$\eta$ Bootis is a V=2.68 mag star of spectral type G0IV extensively studied with
MOST satellite in pursue of its solar-like oscillations \citep{gue07}. 
It exhibits no variability above 0.0001 mag. 
Using PST2G we obtained 173 frames and reduced them using standard
photometric routines with Starlink package scripts,
including correction for bias, dark current and flat field.
Differential aperture photometry of the first order diffraction image of
$\eta$ Boo and 0-order image of the nearby V=7.1 comparison star GSC 1470-0590
yielded standard deviation of individual measurements of 0.026 mag. 
Binning of each 15 measurements, lasting 3.5 minutes,
yields reduced $\chi^2=1.89$ for 10 degrees of freedom and standard deviation
0.009 mag, with respect to a constant. It seems that field rotation
coupled with wire mesh geometry imperfections (as discussed in section \ref{s3})
did not affect photometric results significantly.

\begin{figure}
\begin{center}
\includegraphics[angle=0,scale=.20]{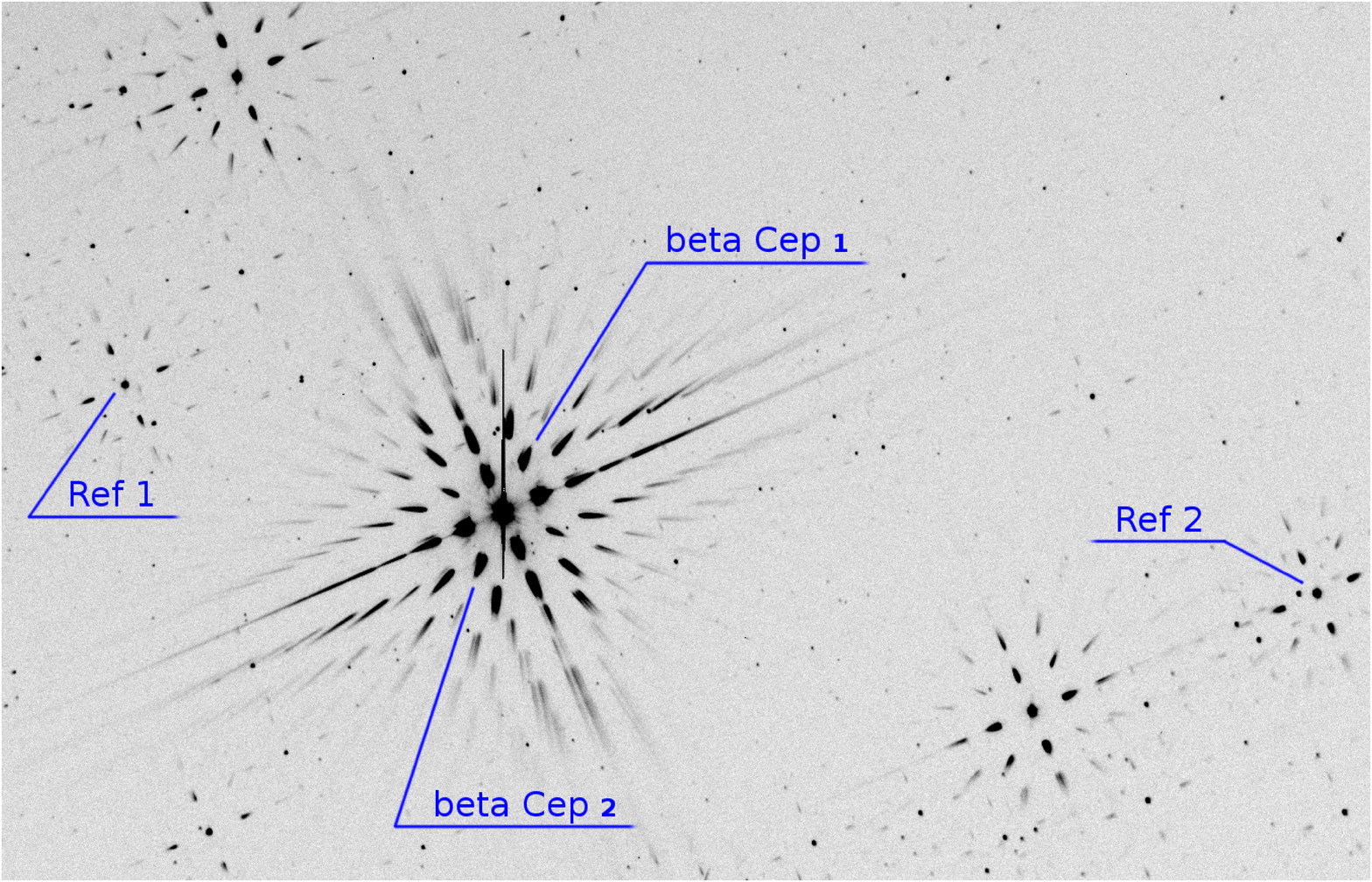}
\caption{Image of $\beta$ Cephei, V$\approx$3.2 mag, taken through the wire mesh. 
Two measured diffraction images of the target star and the reference stars are marked. 
GSC 4465-0882, V=8.2, ($\mbox{Ref 2}$) served as the comparison star, GSC 4465-0481, V=9.0,
($\mbox{Ref 1}$) as a check star.\label{fig1}}
\end{center}
\end{figure}

For further tests we selected as the bright program star $\beta$ Cephei 
(V$\approx$3.2 mag), the archetype of a class of multiperiodic pulsating stars. 
Using the Orion telescope we obtained 150 frames and reduced them the same way
as in the case of $\eta$ Bootis. We used elliptical apertures
to measure the target star first order diffraction images and circular apertures
for central (zero order) images or reference stars
GSC 4465-0481, V=9.0 and GSC 4465-0882, V=8.2 (marked in Fig. \ref{fig1}
as $\mbox{Ref 1}$ and $\mbox{Ref 2}$, respectively).

\begin{figure}
\begin{center}
\includegraphics[angle=270,scale=.60]{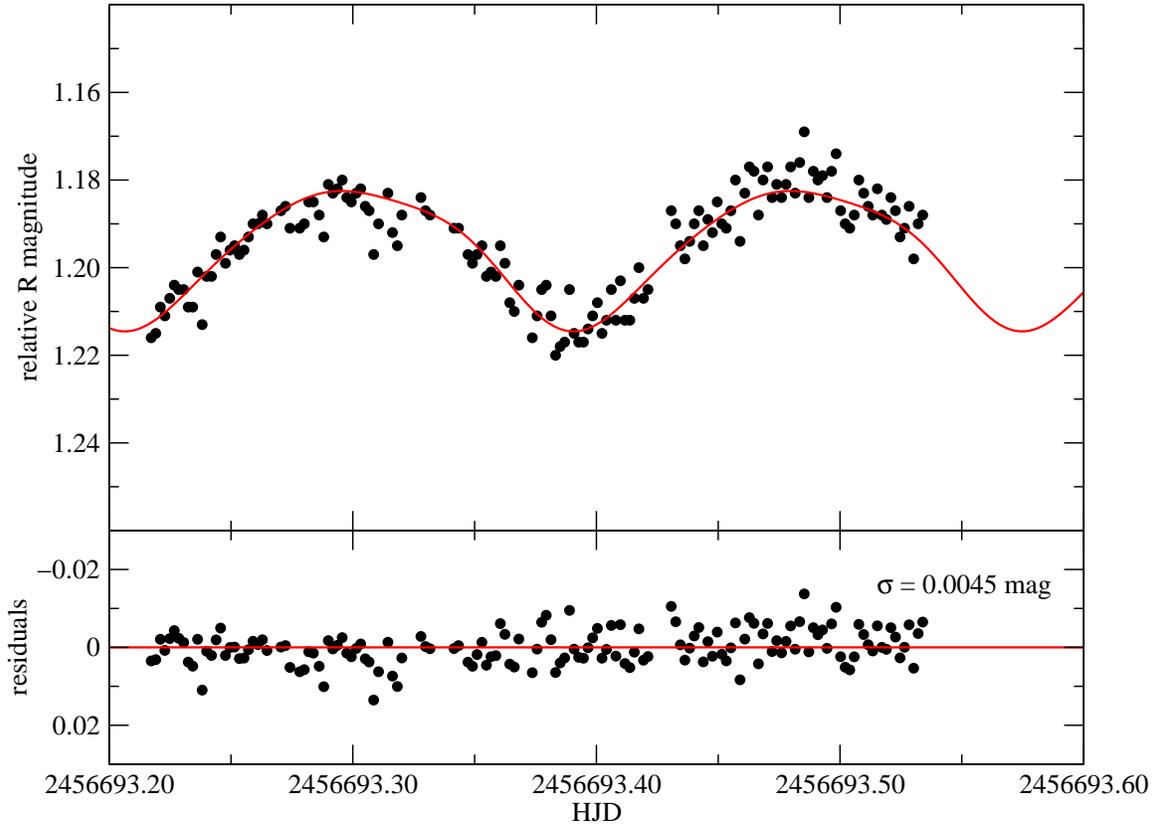}
\caption{Differential light curve of the first order diffraction image of $\beta$ Cephei ($\beta\mbox{~Cep}_1$ in Fig. \ref{fig1})
with respect to 0-order image of the reference star GSC 4465-0882 ($\mbox{Ref 2}$ in Fig. \ref{fig1}).
Each point represents 150s exposure.
Smooth curve marks the least-square fit of 3 harmonic Fourier series. \label{fig2}}
\end{center}
\end{figure}

In Fig. \ref{fig4} we plot magnitude difference $\mbox{Ref 1}-\mbox{Ref 2}$
covering an interval of about 6 hours.
No trend larger than the unweighted standard deviation
of 0.006 mag is present. In Fig. \ref{fig2} we plot 
magnitude difference $\beta\mbox{~Cep}_1-\mbox{Ref 2}$.
To derive an external error estimate
we fitted data with Fourier series of 3 harmonics 
and for 141 degrees of freedom we obtained reduced 
$\chi^2=1.16$. The unweighted standard deviation 
in the plot is 0.0045 mag.
Since the comparison star is redder than target star 
($(B-V)_{Ref 2} = +0.22$, $(B-V)_{\beta Cep} = -0.22$) 
and airmass was growing we attribute a slight linear trend in residuals
at the level of 0.6 mmag per hour to differential extinction.

\begin{figure}
\begin{center}
\includegraphics[angle=270,scale=.60]{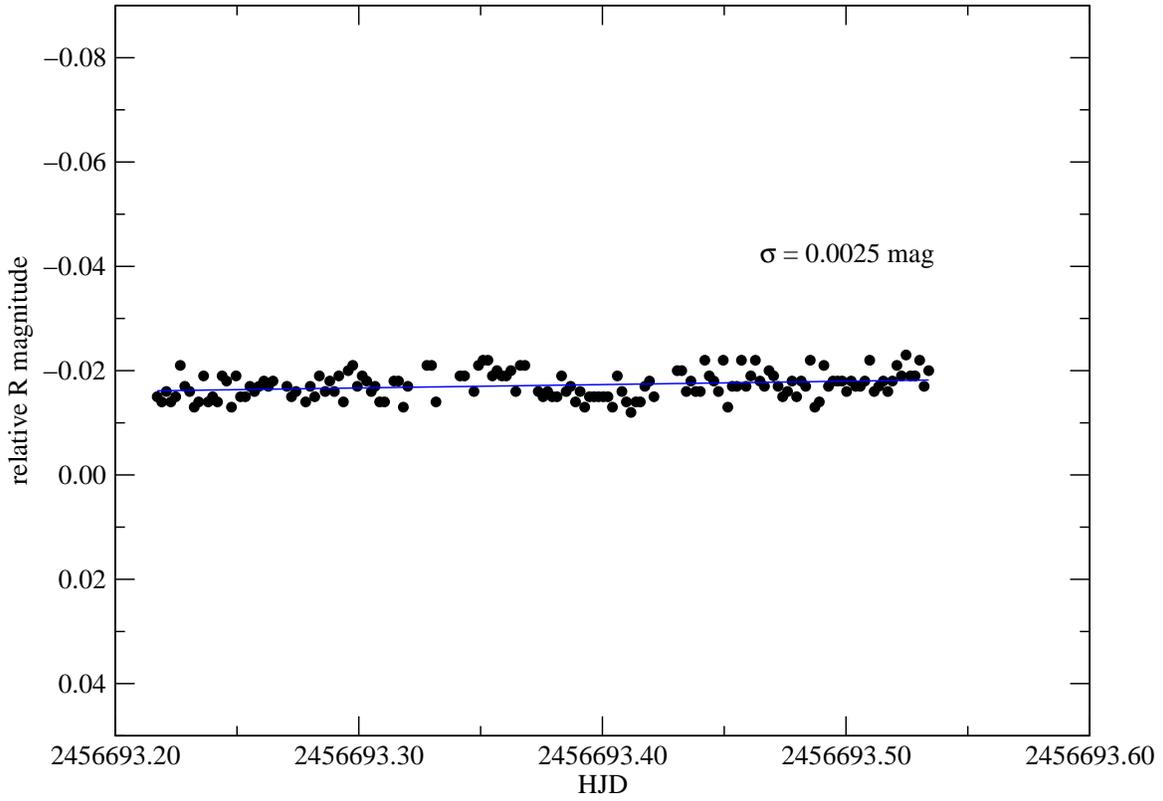}
\caption{Differential photometry between the first order diffraction images of $\beta$ Cephei ($\beta 
\mbox{~Cep}_1 -\beta\mbox{~Cep}_2$). Each 150s exposure is plotted.
The line represents a least-square linear fit. We attribute the non-zero difference
between symmetric diffraction images mostly to the wire mesh geometrical
imperfections.\label{fig3}}
\end{center}
\end{figure}

\begin{figure}
\begin{center}
\includegraphics[angle=270,scale=.60]{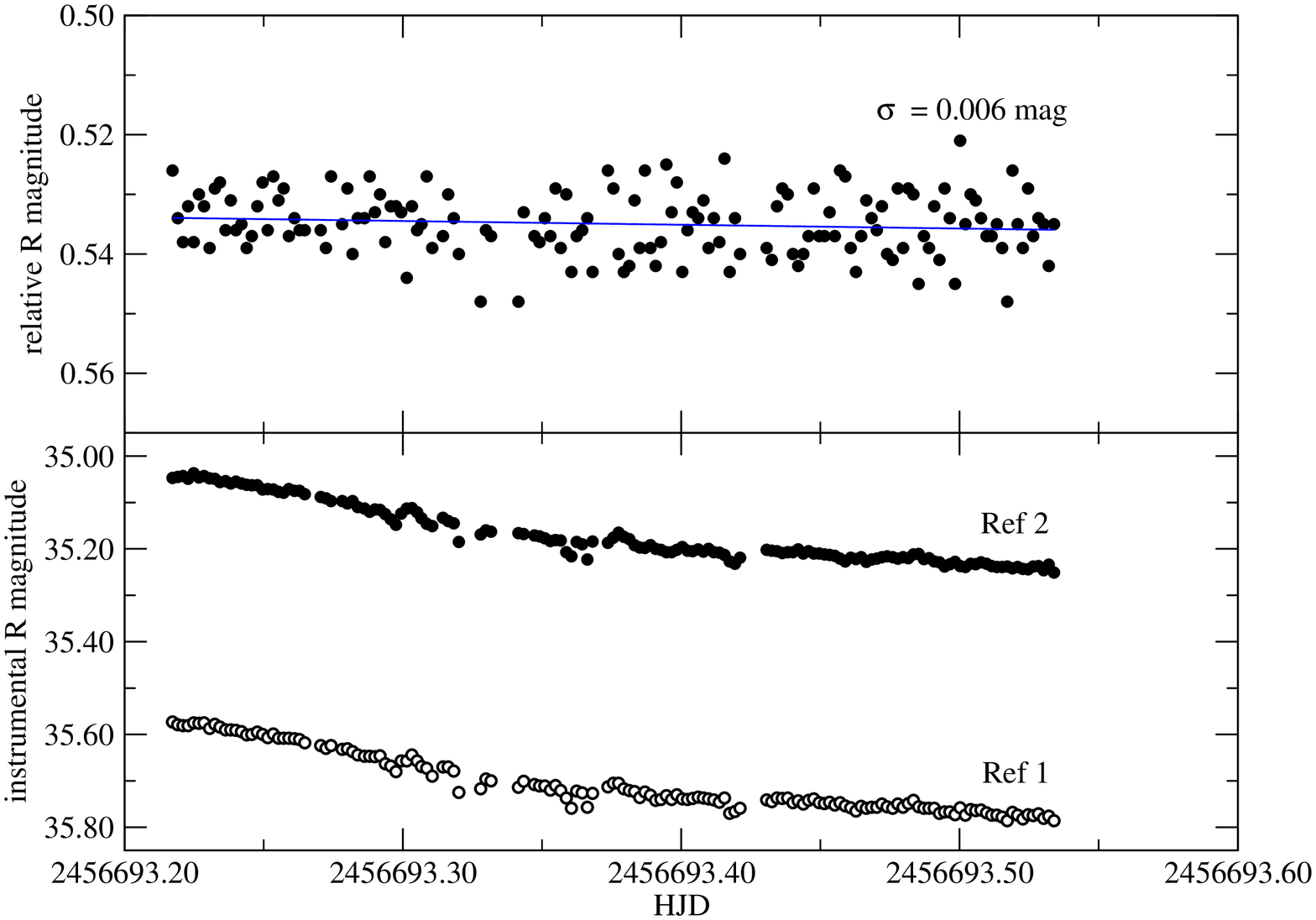}
\caption{Top: Differential photometry between reference stars $\mbox{Ref 1} -\mbox{Ref 2}$.
The line represents a least-square linear fit. Bottom: Instrumental magnitudes of both reference stars.\label{fig4}}
\end{center}
\end{figure}

Inspection of Fig. \ref{fig1} and Fig. \ref{fig3} reveals that diffraction
images intensity and geometry are distorted in different ways, reflecting imperfect
geometry of the wire grid. The two geometries are related by Fourier 
transform, hence the image intensity ratio remains fixed for a fixed mesh 
pattern. For proper aperture centering, the shift of magnitudes between diffraction
images remains fixed too. In particular, the long-scale translation and wire 
thickness asymmetry yield a constant magnitude shift between brightness of 
diffraction images plotted in Fig. \ref{fig3}.

Errors in Figs. \ref{fig2}-\ref{fig4} appear consistent with independent white noise.
Namely, application of additivity of error squares to Fig. \ref{fig3}
yields the standard error of a single measurement of $\beta$ Cep as 
$\sigma_{\beta}= 0.0025/\sqrt{2}= 0.0018$ mag.
If so, then from Fig. \ref{fig2} the error of the comparison star is $\sigma_c= 
\sqrt{0.0045^2-0.0018^2}= 0.0041$, consistent with an independent estimate 
from Fig. \ref{fig4} $\sigma_c= 0.006/\sqrt{2}= 0.0042$.

To evaluate the effect of variable seeing, which should affect each diffraction
order differently, we compared the first and the second diffraction image of the target star.
For two diffraction images of $\beta$ Cep marked in Fig. \ref{fig1}
the standard deviation of magnitude difference does not exceed 0.0025 mag 
and neither do any trends (Fig. \ref{fig3}). 
Thus changes in seeing affect our results by no more than 0.0025 mag. 
Similar results are obtained for other combinations of pairs selected from 
diffraction images close to the center, 
so we conclude that our diffraction image photometry 
remains little affected by variable seeing.

\section{Conclusions} \label{s3}
Several approaches have been utilized in the past for obtaining accurate photometry
of bright stars using CCD detectors, including:
\begin{enumerate}
\item alternate long and short exposures - prone to residual bulk image on CCD
chip and atmospheric condition changes.
Additionally short exposure times are heavily affected by scintillation;
\item snapshot observation technique \citep{man11} - requires precise and
multiple telescope slews and is sensitive to atmospheric condition changes.
Both (1) and (2) require photometric conditions;
\item covering a fraction of the detector with a neutral density filter
reduces useful telescope field of view by introducing a "penumbra" area
and for a given filter yields limited dynamic range.
\end{enumerate}

The objective wire mesh technique described in Sect. \ref{s1} suffers 
from none of these drawbacks and produces useful CCD dynamic range between the 
target and the comparison star of up to at least 5 magnitudes, depending on selection 
of an appropriate fringe of the target star. Even a wider range of magnitude 
differences should be available by thinning of mesh wires, so that the low-order fringes 
become fainter. Thus our technique, combined with the appropriate exposure time, 
permits free choice of the comparison star to meet such criteria like 
scintillation time averaging or appropriate filling of the CCD pixel  well. 
Results of Sect. \ref{s2} demonstrate that in this way excellent photometric 
precision may be reached in poor climate with inexpensive equipment. Immediate 
application of our technique would be for ground follow-up observations for 
BRITE constellation of satellites. Space photometry may reach 
several orders of magnitude better precision than possible 
from the ground. However, due to reliance on mechanical 
devices for accurate pointing its time span is limited and so is 
frequency resolution. Thus, for sufficiently large amplitude oscillations, 
ground observations still remain useful.

The wire mesh does not have to be installed on the objective. It may be 
convenient to place it directly on the photometric filter. In that case 
the mesh cell size should be reduced proportionally to the $a/f$ ratio, where 
$f$ is the telescope effective focal length and $a$ is the distance between the 
mesh and the image. With an internal wire mesh each star fringe pattern 
is created by a different mesh section, but with a proper telescope tracking 
and non-rotating field of view this should always be the same section for a given star.
Therefore, the relative intensity of diffraction fringes is preserved, 
making relative photometry still possible, but differs from star to star, 
which prevents from accurate photometric calibration.
Our test of this variant of the wire mesh technique seems encouraging,
but this concept requires further investigation.

The wire mesh technique could be useful not just for BRITE 
follow-up observations, but could also provide parallel photometric observations 
for high resolution spectroscopic observations with larger telescopes, 
e.g. similar to our PST2 project.

\acknowledgments

Instrumental \& observational work at Poznan Observatory by K.K. is supported by
Polish NCN grant UMO-2011/01/D/ST9/00427.
Studies of structure and evolution of bright stars by A.S.-C. are funded by  
NCN grant UMO-2011/01/M/ST9/05914.
We also thank the referee for useful comments and suggestions.

\clearpage

\clearpage

\clearpage

\clearpage

\clearpage


\begin{thebibliography}{}
\bibitem[Crawford(1968)]{cra68}
         Crawford, F.C., 1968, {\em Waves}, Berkeley Physics Course,
Vol. 3, N.Y., McGraw-Hill.
\bibitem[Daszyńska et al.(2002)]{das02}        
         Daszyńska-Daszkiewicz, J., Dziembowski, W. A., Pamyatnykh, A. A.,
         Goupil, M.-J., 2002, \aap, 392, 151.
\bibitem[Guenther et al.(2007)]{gue07}         
         Guenther, D. B., Kallinger, T., Reegen, P., Weiss, W. W., Matthews, J. M.,
         Kuschnig, R., Moffat, A. F. J., Rucinski, S. M., Sasselov, D., Walker, G. A. H.,
         2007, Communications in Asteroseismology, 151, 5.
\bibitem[Hertzsprung(1910)]{her10}
         Hertzsprung, E., 1910, AN, 184, 237 (in German).
\bibitem[Mann et al.(2011)]{man11}     
         Mann, A. W., Gaidos, E., Aldering, G., 2011, \pasp, 123, 1273. 
\bibitem[Orleanski et al.(2010)]{orl10}
         Orleanski, P., Graczyk, R., Rataj, M., Schwarzenberg-Czerny, A.,
         Zawistowski, T. \& Zee, R., 2010, in {\em Photonics Applications in Astronomy,
         Communications, Industry, and High-Energy Physics Experiments}, Romaniuk, R.
         (Ed.), Proceedings of the SPIE, 7745, 77450A.
\bibitem[Weaver(1946)]{wea46}
         Weaver, Harold F., 1946, Popular Astronomy, 54, 287.
\end{thebibliography}
\end{document}